# Stability of the Holographic Description of the Universe

Peng Huang [1,a]    Yong-chang Huang [1, 2, 3]

[1] Institute of Theoretical Physics, Beijing University of Technology, Beijing 100022, China
[2] Kavli Institute for Theoretical Physics China at the Chinese Academy of Sciences, Beijing, 100080, China
[3] CCAST (World Laboratory), Beijing 100080, China

Abstract

We investigate the stability of the holographic description of the universe. By treating the perturbation globally, we discover that this description is stable, which is support for the holographic description of the universe.

[a] e-mail: phuang@emails.bjut.edu.cn

# 1. Introduction

The cosmological constant problem [1] is a longstanding problem in theoretical physics. Evidence from Type SN Ia [2][3], CMB [4] and SDSS [5] shows that the current universe is accelerating, which can be explained by dark energy (a generalization of the cosmological constant). There are various models about dark energy [6~18], and the holographic dark energy (HDE) model [13~18] has an intrinsic advantage over the other models in that it does not need fine-tuning of the parameters or an ad hoc mechanism to cancel the zero-point energy of the vacuum. The key point in the HDE model is that in quantum field theory, due to the limit made by the formation of a black hole, an ultraviolet (UV) cut-off is related to an infrared (IR) cut-off [13]. Thus, if one labels $\rho_D$ as the quantum zero-point energy density caused by an UV cut-off, the total energy in a region of size $L$ should not exceed the mass of a black hole of the same size, that is, $L^3\rho_D \leq LM_p^2$, so one has $\rho_D = 3C^2M_p^2L^{-2}$, here, $C$ is a numerical constant introduced for convenience and $M_p$ is the reduced Planck mass.

If one supposes that there is no interaction between dark energy and matter, an inevitable result is to use the future event horizon for IR cut-off, only by doing this can we deduce the correct equation of state to obtain an accelerated universe. The holographic dark energy model developed from this viewpoint is as follows [15]:

$$\rho_D = 3C^2M_p^2L_E^{-2}, \qquad (1)$$

where $C$ is a positive numerical parameter which is in favor of $C=1$ [14][15], $M_p$ is the reduced Planck mass, $L_E = a(t)r_E(t)$, the definition of $r_E(t)$ is

$$\int_0^{r_E(t)} \frac{dr}{\sqrt{1-kr^2}} = \frac{R_E(t)}{a(t)} = \int_t^\infty \frac{dt}{a(t)},$$ $R_E(t)$ is the future event horizon, $k =1$, 0, -1 corresponds to closed, flat and open universe, respectively. Though it is well in

agreement with cosmological observation [19], using the event horizon for the large scale cut-off in this HDE model is thought to cause causality problem, see [20] for further reading. Recently, Li *et al* deduce this HDE model from action principle [21], thus, as they claimed, solved the causality problem.

HDE model can be understood in terms of the holographic principle [22-24]. Such a basic principle should be universal and is not only for a special object. Thus, a natural generalization from the HDE model is that the remnant kinds of energy in the universe also have their holographic characters. Efforts along this consideration leads to the establishment of the holographic $E_{KMR}$ model ($E_{KMR}$ is a short expression for energy coming from spatial curvature, matter and radiation together) [25]. In this model,

$$\rho_{KMR} = 3C^2 M_p^2 L_p^{-2}, \tag{2}$$

where $\rho_{KMR} = \rho_K + \rho_M + \rho_R$, $C$, and $M_p$ have the same meaning as that in Eq.(1), $L_p = a(t) r_p(t)$, $r_p(t)$ is defined by $\int_0^{r_p(t)} \frac{dr}{\sqrt{1-kr^2}} = \frac{R_p(t)}{a(t)} = \int_0^t \frac{dt}{a(t)}$, $R_p(t)$ is the particle horizon, $k=1$ corresponds to the closed universe.

The holographic dark energy together with the holographic $E_{KMR}$ give a holographic description of the closed and accelerated expanding universe [25]. As a phenomenological model, the stability of the holographic description of the universe is an important problem. Li *et al* have calculated the stability of *HDE* in flat universe [26] in which the perturbations coming from radiation and matter are neglected, it is all right when the attention is just focused on the *HDE*, but when one study the stability of the holographic universe consisted of *HDE* and $HE_{KMR}$, perturbations coming from dark energy, spacial curvature, matter and radiation all must be taken into consideration. We will investigate these in detail in this paper.

The plan of the paper is as follows. In Sec. II we get the variation of the energy density; in Sec. III we give detailed discussions of the stability; the last section is summary and conclusion.

## 2. The variation of the energy density

In the previous section, we have labeled the energy densities of dark energy, matter, radiation and the energy coming from spatial curvature as $\rho_D$, $\rho_M$, $\rho_R$ and $\rho_K$ respectively, the Friedmann equations says

$$3M_p^2 H^2 = \rho - \frac{3kM_p^2}{a(t)^2}, \tag{3}$$

which is just

$$\rho_c = \rho_D + \rho_M + \rho_R - \rho_K, \tag{4}$$

where $\rho_c = 3M_p^2 H^2$ is the critical density and $\rho_K = \frac{3kM_p^2}{a(t)^2}$ is the energy density coming from spatial curvature. Equation (4) can be rewritten as

$$\rho_c + 2\rho_K = \rho_D + \rho_{KMR}, \tag{5}$$

where $\rho_{KMR} = \rho_K + \rho_M + \rho_R$ is used, so we have

$$\delta(\rho_D + \rho_{KMR}) = \delta(\rho_c + 2\rho_K), \tag{6}$$

inserting $\rho_c = 3M_p^2 H^2$, $\rho_K = \frac{3kM_p^2}{a^2}$ and $H = \frac{\dot{a}}{a}$ into Eq.(6), we have

$$\delta(\rho_D + \rho_{KMR}) = 6M_p^2 H \frac{d}{dt}\frac{\delta a}{a} - 12kM_p^2 \frac{\delta a}{a^3}, \tag{7}$$

so, if $\delta a$ is given, we can get the variation of the energy.

Now, let us study what $\delta a$ is. For simplicity, we focus our attention on the close universe with $k = 1$, the results for the case of an open universe with $k = -1$ can be obtained from those for a closed universe by a transformation.

We consider the perturbation of a scalar type of the metric. The perturbed metric in the Newtonian gauge is [27]

$$ds^2 = -(1 + 2\Phi(r,t))dt^2 + a^2(1 - 2\Phi(r,t))(\frac{dr^2}{1-r^2} + r^2 d\Omega^2), \tag{8}$$

where, as usual, the perturbation is spherically symmetric. Thus, the future event horizon $R_E$ can be expressed as

$$R_E(t) = a \int_0^{r_E(t)} (1 - \Phi(r,t)) \frac{dr}{\sqrt{1-r^2}}. \tag{9}$$

When there is no perturbation, the future event horizon $R_{E0}$ is

$$R_{E0}(t) = a \int_0^{r_{E0}(t)} \frac{dr}{\sqrt{1-r^2}}. \tag{10}$$

So, the variation of the event horizon $R_{E0}$ can be defined as

$$\delta R_{E0}(t) = R_E(t) - R_{E0}(t) = a[\sin^{-1} r_E - \sin^{-1} r_{E0}] - a \int_0^{r_{E0}(t)} \Phi(r,t) \frac{dr}{\sqrt{1-r^2}}, \tag{11}$$

which is just

$$\delta R_{E0}(t) = R_E(t) - R_{E0}(t) = a\delta \sin^{-1} r_{E0} - a \int_0^{r_{E0}(t)} \Phi(r,t) \frac{dr}{\sqrt{1-r^2}}$$

$$= \frac{a}{\sqrt{1-r_{E0}^2}} \delta r_{E0} - a \int_0^{r_{E0}(t)} \Phi(r,t) \frac{dr}{\sqrt{1-r^2}}. \tag{12}$$

On the other hand, we know from $\int_0^{r_{E0}(t)} \frac{dr}{\sqrt{1-r^2}} = \frac{R_{E0}(t)}{a}$ that

$R_{E0}(t) = a \sin^{-1} r_{E0}(t)$, which tells

$$\delta R_{E0}(t) = a\delta \sin^{-1} r_{E0}(t) + \sin^{-1} r_{E0}(t) \delta a$$

$$= \frac{a}{\sqrt{1-r_{E0}^2}} \delta r_{E0} + \sin^{-1} r_{E0}(t) \delta a. \tag{13}$$

Comparing Eqs. (12) and (13), we find

$$\delta a = -\frac{a}{\sin^{-1} r_{E0}} \int_0^{r_{E0}} \Phi(r,t) \frac{dr}{\sqrt{1-r^2}}. \tag{14}$$

Inserting Eq.(14) into Eq.(7), we can have

$$\delta(\rho_D + \rho_{KMR}) = -\frac{6M_p^2 H}{\sin^{-1} r_{E0}} \frac{d}{dt} \int_0^{r_{E0}(t)} \Phi \frac{dr}{\sqrt{1-r^2}} + \frac{6M_p^2 H \dot{r}_{E0}}{(\sin^{-1} r_{E0})^2 \sqrt{1-r_{E0}^2}} \int_0^{r_{E0}} \Phi \frac{dr}{\sqrt{1-r^2}}$$

$$+\frac{12M_p^2}{a^2\sin^{-1}r_{E0}}\int_0^{r_{E0}}\Phi(r,t)\frac{dr}{\sqrt{1-r^2}}$$

$$=-\frac{6M_p^2 H\Phi(r_{E0},t)\dot{r}_{E0}}{\sin^{-1}r_{E0}\sqrt{1-r_{E0}^2}}+\frac{6M_p^2 H\dot{r}_{E0}}{(\sin^{-1}r_{E0})^2\sqrt{1-r_{E0}^2}}\int_0^{r_{E0}}\Phi(r,t)\frac{dr}{\sqrt{1-r^2}}$$

$$+\frac{12M_p^2}{a^2\sin^{-1}r_{E0}}\int_0^{r_{E0}}\Phi(r,t)\frac{dr}{\sqrt{1-r^2}}, \qquad (15)$$

and we can derive from $\int_0^{r_{E0}(t)}\frac{dr}{\sqrt{1-r^2}}=\int_t^\infty\frac{dt}{a(t)}$ that $\dot{r}_{E0}(t)=-\frac{\sqrt{1-r_{E0}^2}}{a}$, inserting this into Eq.(15), finally, we get the variation of the energy density as follows:

$$\delta(\rho_D+\rho_{KMR})=\frac{6M_p^2 H\Phi(r_{E0},t)}{R_{E0}(t)}+(\frac{12M_p^2}{a^2\sin^{-1}r_{E0}}-\frac{6M_p^2 H}{R_{E0}(t)\sin^{-1}r_{E0}})\int_0^{r_{E0}}\Phi(r,t)\frac{dr}{\sqrt{1-r^2}} \qquad (16)$$

Once we know the variation of the energy density of the holographic universe, we can learn its stability through the perturbative Einstein equation. We study this in detail in the next section.

## 3. The stability of the holographic description of the universe

Using the 00-component of the perturbative Einstein equation ( i.e. the first equation in Eq.(4.15) of Ref. [28]) in the Newtonian gauge, one has

$$\frac{\nabla^2}{a^2}\Phi-3H\dot{\Phi}-3H^2\Phi+\frac{3\Phi}{a^2}=\frac{1}{2M_p^2}(\delta\rho_D+\delta\rho_{KMR}), \qquad (17)$$

using Eq.(16), we obtain the concrete formalism of the perturbative Einstein equation

$$\frac{\nabla^2}{a^2}\Phi-3H\dot{\Phi}-3H^2\Phi+\frac{3\Phi}{a^2}$$

$$=-3H[(\frac{1}{R_{E0}\sin^{-1}r_{E0}}-\frac{2}{R_{E0}Ha})\int_0^{r_{E0}(t)}\Phi(r,t)\frac{dr}{\sqrt{1-r^2}}-\frac{\Phi(r_{E0},t)}{R_{E0}}]. \qquad (18)$$

In order to solve this equation, analogous to Ref.[26]'s discussion, we expand $\Phi$ as $\Phi(r,t)=\sum_l\Phi_l(t)\frac{\sin lr}{r}$ where we have dropped the $\frac{\cos lr}{r}$ terms which causes a singularity when r = 0, then, Eq.(18) can be simplified as

$$\frac{\sin lr}{r}[\frac{l^2}{a^2}\Phi_l + 3H\dot{\Phi}_l + 3H^2\Phi_l - \frac{3}{a^2}\Phi_l]$$

$$= 3H[(\frac{1}{R_{E0}\sin^{-1}r_{E0}} - \frac{2}{R_{E0}Ha})\Phi_K(t)\int_0^{r_{E0}(t)}\frac{\sin lr}{r}\frac{dr}{\sqrt{1-r^2}} - \frac{\Phi_l(t)}{R_{E0}}\frac{\sin lr_{E0}}{r_{E0}}], \quad (19)$$

where we also neglect $\frac{\cos lr}{r^2}$ and $\frac{\sin lr}{r^3}$ terms of causing singularities when r = 0.

What we concern about here is the behavior of $\frac{\dot{\Phi}_l}{\Phi_l}$. There are two cases relative to the stability: (i) The perturbation mode is frozen if $\frac{\dot{\Phi}_l}{\Phi_l} \to 0$; (ii) the perturbation mode is decaying if $\frac{\dot{\Phi}_l}{\Phi_l} < 0$. Otherwise, the holographic description of the universe is unstable.

We first study the super-horizon mode where $lr_{E0}(t) \ll 1$. In this case, Eq.(19) can be rewritten as

$$3H\dot{\Phi}_l + 3H^2\Phi_l - \frac{3}{a^2}\Phi_l = 3H[(\frac{1}{R_{E0}\sin^{-1}r_{E0}} - \frac{2}{R_{E0}Ha})\Phi_l(t)\sin^{-1}r_{E0} - \frac{\Phi_l(t)}{R_{E0}}], \quad (20)$$

which is just

$$3H\dot{\Phi}_l(t) + 3H^2\Phi_l(t) - \frac{3}{a^2}\Phi_l(t) = -\frac{6}{a^2}\Phi_l(t). \quad (21)$$

It follows from Eq.(21) that

$$\frac{\dot{\Phi}_l}{\Phi_l} = -\frac{1}{H}(\frac{1}{a^2} + H^2), \quad (22)$$

which is definitely negative. So the super-horizon mode is a decaying mode, there is no instability for the perturbation.

For the sub-horizon mode $lr_{E0}(t) \gg 1$, the integration in the RHS of Eq.(19) is an oscillating term whose value can be regarded as zero; thus, the the dominant contribution comes from $-3H\frac{\Phi_l(t)}{R_{E0}}\frac{\sin lr_{E0}}{r_{E0}}$; the dominant contribution in the LHS of Eq.(19) comes from $\frac{\sin lr_{E0}}{r_{E0}}[\frac{l^2}{a^2}\Phi_l + 3H\dot{\Phi}_l + 3H^2\Phi_l - \frac{3}{a^2}\Phi_l]$, therefore, Eq.(19) turns to

$$\frac{l^2}{a^2}\Phi_l(t) + 3H\dot{\Phi}_l(t) + 3H^2\Phi_l(t) - \frac{3\Phi_l(t)}{a^2} + \frac{3H\Phi_l(t)}{R_{E0}} = 0, \quad (23)$$

which can be reformed to

$$\frac{\dot{\Phi}_l}{\Phi_l} = -\frac{1}{3H}(\frac{l^2-3}{a^2} + 3H^2 + \frac{3H}{R_{E0}}), \quad (24)$$

which is also definitely negative, again, no instability appears.

So, we can see that the holographic universe is stable under perturbation. This is support for the holographic description of the universe.

The universe we study above is a closed one. For observational reasons, the question of whether these holographic energy models are stable in a flat universe is also in need of an answer. In fact, when $k=0$, Ref.[14] has proven that,

$$\frac{\Omega'_D}{\Omega_D^2} = (1-\Omega_D)(\frac{1}{\Omega_D} + \frac{2}{c\sqrt{\Omega_D}}), \quad (25)$$

with $\Omega_D = \frac{\rho_D}{\rho_c}$ and the prime denotes the derivative with respect to $\ln a$, which says that $\Omega'_D$ is always positive and thus dark energy will be dominant in the universe. Because of the dominant character of the dark energy in the future universe, the problem mentioned above turns into a new one: whether the holographic dark energy model is stable in flat universe? After similar calculation, one can find that the behavior of the super-horizon and sub-horizon modes are described as follows [26], super-horizon: $\quad \frac{\dot{\Phi}_l}{\Phi_l} = \frac{H(1-\rho_D/\rho_c)}{(2C\sqrt{\rho_D/\rho_c}-1)} < 0, \quad (26)$

sub-horizon: $\quad \frac{\dot{\Phi}_l}{\Phi_l} \approx -\frac{(l^2/a^2 + 3H^2)}{3H} < 0. \quad (27)$

These results assure the stability of the HDE model in flat universe which is important for cosmology.

## 4. Summary and Conclusion

In this paper, we study the stability of the holographic description of the universe. We first get the relationship between $\delta(\rho_D + \rho_{KMR})$ and $\delta a$ from the Friedmann equations, then we derive the value of $\delta a$ from the perturbative metric in the Newtonian gauge. After these two steps we immediately obtain the concrete formalism of perturbative Einstein equation, we discuss the stability on the basis of the perturbative equation and find that the holographic description of the universe is stable.

It needs to point out that there is another way to investigate the stability. For a usual fluid, we can calculate the square of the speed of sound in the usual fluid and find out whether it is negative or not; the negative case leads to instability and the positive case results in the opposite side. It seems that we can assume that the holographic universe is filled up with a usual fluid, then we can do the calculation mentioned above. But for holographic $E_{KMR}$ and holographic dark energy, the perturbation of the energy density is nonlocal, it comes from the perturbation of the metric which must be treated globally; for a similar reason as has been pointed out in Ref [26]. This is completely different from a usual fluid component, and does not suffer such instability.

The motivation to study the stability of the holographic description of the universe is that the stability is an important issue for phenomenological cosmological models; only when such models is stable under perturbation can we have the opinion that maybe it is on the correct way to describe the universe. A technical difference between the investigation of the stability of the holographic dark energy model and the stability of the holographic description of the universe is that the use of the Friedmann equation, which can be seen from Eq.(3) to Eq.(6), variation of the energy density carried out in this way makes the calculation of the perturbated Einstein equation easy significantly; if one gets the variation of the energy density in other ways, such as denoting the variation of the energy density coming from $HDE$ and $HE_{KMR}$ with respect to the metric perturbation, respectively, the treatment of the perturbed Einstein equation will become very difficult. The novel feature in our treatment is that, perturbations coming from matter, radiation and spacial curvature are always neglected when one focus just on the holographic dark energy model; but in our investigating of the stability of the holographic universe consisted of $HDE$ and $HE_{KMR}$, perturbations coming from spacial curvature, matter and radiation all are taken into consideration.